\documentclass[3p,times,twocolumn]{elsarticle}

\usepackage{ecrc}
\usepackage{braket}
\usepackage[english]{babel}

\volume{00}

\firstpage{1}

\journalname{Nuclear Physics B Proceedings Supplement}

\runauth{E. Czerwi\'nski}

\jid{nuphbp}

\jnltitlelogo{Nuclear Physics B Proceedings Supplement}

\usepackage{amssymb}
\usepackage{amsmath}

 \biboptions{sort&compress}

\usepackage[figuresright]{rotating}

\def\raE     {\rightarrow}

\def \Kp3pi  {${\rm K^+} \rightarrow \pi^+\pi^-\pi^+$~}
\def \Kp3pig {${\rm K^+} \rightarrow \pi^+\pi^-\pi^+(\gamma)$~}
\def \Kmunu  {${\rm K^{\pm}} \rightarrow \mu^{\pm}\nu(\gamma)$~}
\def \Kpipi  {${\rm K^{\pm}} \rightarrow \pi^{\pm}\pi^0(\gamma)$~}
\begin{document}

\begin{frontmatter}



\dochead{}

\title{Recent KLOE results on kaon physics}


\author{E. Czerwi\'nski on behalf of KLOE--2 Collaboration}

\address{Institute of Physics, Jagiellonian University, 30-059 Cracow, Poland}

\begin{abstract}
A short review of two measurements recently published by KLOE/KLOE--2 collaborations is presented.
Namely the most stringent limits of the CPT-violating parameters $\Delta a_\mu$ for neutral kaons, and
the absolute branching ratio of $K^+\to\pi^+\pi^-\pi^+(\gamma)$ decay.
In addition a discussion of starting KLOE--2 project is performed.
\end{abstract}


\end{frontmatter}


\section{Introduction}
The K LOng Experiment completed its data taking campaign in 2006.
The detector designed for neutral and charged kaons studies operated at the DA$\Phi$NE $e^+e^-$ collider~\cite{Vignola:1996mt}
with a c.m. energy working point equal to the $\phi$ meson mass,
where kaons were produced in pairs from the $\phi$ meson decay (BR($\phi\to\ KK)\approx 83\%$).
Total data sample corresponding to 2.5fb$^{-1}$ of integrated luminosity
allowed to study discrete symmetries and to precisely measure dominant kaon branching ratios. 
Recently an upgraded KLOE detector started operation to extend KLOE physics program (KLOE--2).
\section{KLOE detection setup}
The KLOE system consists mainly of two detectors: a cylindrical drift chamber with diameter of 4m~\cite{Adinolfi:2002uk} surrounded by an electromagnetic
calorimeter~\cite{Adinolfi:2002zx},
both immersed in
0.5T magnetic field. High performance of drift chamber for momentum and vertex reconstruction
(${\sigma_{p_{\perp}}}/{p_{\perp}}<0.4\%$ for
$\theta>45^{\circ}$;
$\sim$150~$\mu$m in transverse plane)
and excellent time and energy resolution of the calorimeter
($\sigma_{t}=57ps/\sqrt{E(GeV)}\oplus100ps$;
$\sigma_{E}/E=5.7\%/\sqrt{E(GeV)}$)
ensure high quality of collected data. The $\phi$ meson produced in $e^+e^-$ collisions has only a small momentum in $z$ direction ($\sim15$MeV). Since
kaons are always produced in pairs from $\phi$ decay, identification of only one of them allows for four-momentum determination of the second one.
This is the so called tagging technique.
\section{CPT symmetry and Lorentz invariance test}
The CPT symmetry as a simultaneous composition of charged conjugation (C), parity (P) and time reversal (T)
appears to be the only respected discrete symmetry in Nature.
Tiny CPT symmetry violations might appear in conjunction with Lorentz symmetry breaking~\cite{Greenberg:2002uu}.
In view of an effective field theory (Standard Model Extension)~\cite{Kostelecky:1994rn,Colladay:1996iz,Colladay:1998fq}
for neutral kaons
the CPT violation is introduced in the mixing parameter $\delta_K$, with an additional dependence on the four-momentum of kaon:
\begin{eqnarray}
\delta_K\approx i~\mathrm{sin}\phi_{SW}\mathrm{e}^{i\phi_{SW}}\gamma_K(\Delta a_0-\vec{\beta}_K\cdot\Delta\vec{a})/\Delta m,
\end{eqnarray}
where $\gamma_K$ and $\vec{\beta}_K$ are the boost factor and velocity of the kaon
in the observer rest frame, respectively, $\phi_{SW}=\mathrm{arctan}(2\Delta m\slash\Delta\Gamma)$
is the superweak phase with $\Delta m$ and $\Delta\Gamma$ the differences of mass and width between $K_S$ and $K_L$, respectively, and $\Delta a_{\mu}$
are four $CPT$ and Lorentz violating coefficients~\cite{Kostelecky:1994rn,Colladay:1996iz,Colladay:1998fq}.
Determination of these parameters are naturally performed in the reference frame of fixed stars.
Since $\phi$ meson is produced at DA$\Phi$NE almost at rest 
$p_1\sim-p_2$ and
$\delta_K(\vec{p}_1)\ne\delta_K(\vec{p}_2)$,
where $\vec{p}_i$ denotes momentum of each kaon.
On the other side the K mesons produced in $\phi$ meson decays are in a coherent quantum state
($J^{PC}=1^{--}$) and the initial state can be written as:
\begin{eqnarray}
\begin{split}
\left|i\right>=\frac{N}{\sqrt{2}}\left[\left|K_S(\vec{p}_1)\right>\left|K_L(\vec{p}_2)\right> \right. \\
\left. -\left|K_L(\vec{p}_1)\right>\left|K_S(\vec{p}_2)\right>\right],
\end{split}
\label{eq:initial}
\end{eqnarray}
where
$N={\sqrt{\left(1+|\epsilon_S|^2\right)\left(1+|\epsilon_L|^2\right)}}\slash({1-\epsilon_S \epsilon_L})\approx 1$ is a normalization factor,
and  $\epsilon_{S,L}=\epsilon_K\pm\delta_K$. The known contribution from $CP$ symmetry violation is introduced as $\epsilon_K$.
The experimental observable from the integration of equation (\ref{eq:initial})
over the sum of proper decay times
$\tau_1+\tau_2$ at fixed time difference $\Delta\tau=\tau_1-\tau_2$ is the following~\cite{handbook}:
\begin{eqnarray}
\begin{split}
I_{f_1f_2}(\Delta\tau)\propto\mathrm{e}^{-\Gamma\left|\Delta\tau\right|}\left[ \left| \eta_1\right|^2 e^{\frac{1}{2}\Delta\Gamma\Delta\tau}
+\left| \eta_2\right|^2 e^{\frac{1}{2}\Delta\Gamma\Delta\tau} \right.  \\
\left.  -2\mathrm{Re}\left(\eta_1\eta_2^*\mathrm{e}^{-i\Delta m\Delta\tau} \right)\right]
\end{split}
\label{eq:exp}
\end{eqnarray}
where
{$\eta_j=\bra {f_j}T\ket{K_L}\slash\bra{f_j}T\ket{K_S}\simeq\epsilon_K-\delta_K(\vec{p}_j,t_s)$},
$f_1$ and $f_2$ denote kaon final states, $\Gamma=\Gamma_S+\Gamma_L$.
In the reported measurement $f_1=f_2=\pi^+\pi^-$
and due to the fully destructive quantum interference at $\Delta\tau=0$ the distribution (\ref{eq:exp}) is sensitive to changes of $\eta_1\slash\eta_2$ ratio.

At KLOE the measurement of $\phi\to K_SK_L\to\pi^+\pi^-\pi^+\pi^-$ reaction has been performed in order to obtain the $\Delta a_\mu$ parameters from 
the fit of equation (\ref{eq:exp}) to experimental data.
The selection criteria are based on:
\begin{itemize}
\item
the invariant mass of the tracks  connected to a vertex,
\item
two-body kinematics,
\item
missing momentum and \mbox{energy}.
\end{itemize}
The total contamination of signal is 1.5\% mostly from kaons regenerated on the beam pipe,
whereas the ave\-ra\-ge signal efficiency is $\sim$25\%.
Taking into account detector localization on Earth as well as rotation and movement of Earth itself
the space-time coordinates of selected events are transformed to the sidereal frame.
A proper fit of the experimental $I(\Delta\tau)$ distribution yields the following values of the four parameters of Standard Model Extension at KLOE are~\cite{Babusci:2013gda}:
\begin{center}
$\Delta a_0 = (-6.0~\pm~7.7_{stat}~\pm~3.1_{syst})\times 10^{-18}$~GeV,\\
$\Delta a_x = (\hphantom{-}0.9~\pm~1.5_{stat}~\pm~0.6_{syst})\times 10^{-18}$~GeV,\\
$\Delta a_y = (-2.0~\pm~1.5_{stat}~\pm~0.5_{syst})\times 10^{-18}$~GeV,\\
$\Delta a_z = (\hphantom{-}3.1~\pm~1.7_{stat}~\pm~0.5_{syst})\times 10^{-18}$~GeV.
\end{center}
At present the reported values are the most precise measurement of these parameters in the quark sector of Standard Model Extension
and the first independent measurement of all four parameters in the kaon sector.

\section{BR($K^+\to\pi^+\pi^-\pi^+(\gamma)$) measurement}
The last measurement of BR($K^{\pm}\rightarrow \pi^{\pm}\pi^+ \pi^-)$ was performed in 1972 without information about the radiation cut-off~\cite{chiang}
whereas value reported in PDG~\cite{PDG}
is obtained from a global fit that does not use any of the available BR($K^{\pm}\rightarrow \pi^{\pm}\pi^+ \pi^-)$ measurements
but the rate measurement $\Gamma(\pi^+\pi^+\pi^-) = (4.511 \pm 0.024)\times 10^6$~s$^{-1}$
published in 1970~\cite{Ford}.
In the $\pi^0\pi^0$ invariant mass distribution of data collected by NA48 a cusp-like anomaly at
$M_{00} = 2m_{\pi^+}$ can be observed~\cite{NA48-1}. The interpretation is the final state charge-exchange reaction
$\pi^+\pi^- \rightarrow \pi^0\pi^0$ in $K^{\pm} \rightarrow \pi^{\pm}\pi^+\pi^-$ decay~\cite{Budini,Cabibbo}.
Based on the fit of models~\cite{Cabibbo-Isidori} and~\cite{Berna1,Berna2} to experimental $M_{00}^2$ distribution
the difference
between the S-wave $\pi\pi$ scattering lengths in the isospin $I$=0 and $I$=2 states was determined~\cite{NA48-2}, where the main source of uncertainty
was due to the ratio of the branching ratios $K^{\pm} \raE \pi^{\pm}\pi^-\pi^+$ and $K^{\pm} \raE \pi^{\pm}\pi^0\pi^0$.

New measurement performed at KLOE~\cite{Babusci:2014hxa} is based 
on the two samples selected with the usage of ~\Kmunu ($K_{\mu2}$ tags) and ~\Kpipi ($K_{\pi2}$ tags) events.
These independent samples of pure kaons for the signal selection 
are useful for systematic uncertainties evaluation and cross-checks~\cite{KpmSemil}.
The above mentioned decays are identified from the momentum of the charged secondary
track in the kaon rest frame evaluated using the pion mass hypothesis,
the selection efficiency of the two tagging normalization samples are similar and about 36\%.
In order to reduce influence of the trigger efficiency on the signal side a normalization sample of $K_{\mu2}$ or $K_{\pi2}$ tags is selected.
In addition $K^-$ is used as the tagging kaon ($K_{\mu2}$ or $K_{\pi2}$) and $K^+$ as the tagged kaon (signal) due to
factor of $\sim 10^3$ lower the nuclear cross section for positive kaons with momenta $\sim$100MeV with
respect to that of negative kaons~\cite{Knuclear}.

To evaluate the momentum of the tagged kaon at the interaction point (IP) the momentum of the tagging kaon at the IP (backward extrapolated from its first hit in the
drift chamber (DC)) and momentum of the $\phi$-meson measured run by run with Bhabha scattering events are used.
Then momentum of $K^+$ is extrapolated inside the DC (path of the signal kaon).
Requirement of the position of $K^+$ decay vertex inside inner radius of the drift chamber is applied in order to decrease number of charged track
in the DC, since both kaon and pions have momenta lower than $\sim200$MeV and therefore  curl up in the KLOE magnetic field, which increases
number of tracks reconstructed with low quality. Additionally only two of the pion tracks are reconstructed to search for a vertex along the signal kaon path.

The number of $K^+\raE \pi^+\pi^-\pi^+(\gamma)$ is extracted from the
comparison of the MC-predicted shapes for the signal and the background and the experimental 
missing mass spectrum $m^2_{miss} = E^2_{miss} - ({p}_{K^+} - {p}_1 - {p}_2)^2$
where ${p}_1$ and ${p}_2$ are the momenta of the selected tracks.

The branching ratio is given by:
\begin{equation}
BR(K^+ \raE \pi^+\pi^-\pi^+(\gamma)) = \frac{N_{K\raE3\pi}}{N_{tag}}\times\frac{1}{\epsilon}
\end{equation}
where $N_{K\raE3\pi}$ is the number of signal events, $N_{tag}$ is the number of tagged events and $\epsilon$ is
the overall signal selection efficiency (detector acceptance, reconstruction efficiency, tag bias, corrections for the machine and cosmic-ray background).

The effect of the influence of the charged kaon lifetime through the detector acceptance on the BR($K^+\rightarrow \pi^+\pi^+ \pi^-(\gamma)$) 
was also evaluated based on the MC simulation and then applied as weight to the MC events, both for the signal and the control sample
selection procedures.

Averaging two results for $K^- \raE \mu^- \bar{\nu} (\gamma)$ and $K^- \raE \pi^- \pi^0 (\gamma)$ samples, accounting for correlations, the final result of 
the measurement is~\cite{Babusci:2014hxa}:
\begin{equation}
\begin{split}
BR(K^+ \raE \pi^+\pi^-\pi^+(\gamma)) = 0.05565 \\
\pm 0.00031_{stat} \pm 0.00025_{syst},
\end{split}
\end{equation}
which is fully inclusive of final-state radiation and has a 0.72$\%$ accuracy, 
which makes it a factor $\simeq$ 5 better with respect to the previous measurement~\cite{chiang}.
\section{KLOE--2 project}
The KLOE detection system is now upgraded with the Inner Tracker detector~\cite{it}, which will improve resolution on the
vertex position and acceptance for tracks with low transverse momentum; with two pairs of small angle tagging devices to detect
low (Low Energy Tagger - LET~\cite{bib:let}) and high (High Energy Tagger - HET~\cite{taggers})
energy $e^{+}e^{-}$ originated from $e^{+}e^{-}\to e^{+}e^{-}X$ reactions; 
new crystal calorimeters (CCALT) to cover the low polar angle
region to increase acceptance for very forward electrons and photons down to 8$^\circ$~\cite{CCALT};
and a tile calorimeter (QCALT) used for the detection of photons coming from $K_L$ decays in the drift chamber~\cite{QCALT}.
These upgrades will improve systematic uncertainties for KLOE--2 measurements~\cite{AmelinoCamelia:2010me},
while statistical uncertainties will be improved due to higher luminosity of
the DA$\Phi$NE collider.
\section{Conclusions}
Recent results of KLOE explore fundamental symmetries with 10$^{-18}$GeV sensitivity on Standard Model Extension parameters
with application of kaon interferometry, and completes the KLOE program of precision measurements of the dominant kaon branching ratios.
The new KLOE--2 setup started operations at an upgraded DA$\Phi$NE collider for an extension of the original KLOE physics program.
\section{Acknowledgments}
This work was supported in part by the EU Integrated Infrastructure Initiative Hadron Physics Project under contract number RII3-CT-2004-506078; by the European Commission under the 7th Framework Programme through the `Research Infrastructures' action of the `Capacities' Programme, Call: FP7-INFRASTRUCTURES-2008-1, Grant Agreement No. 227431; by the Polish National Science Centre through the Grants No.
2011/03/N/ST2/02641,
2011/01/D/ST2/00748,
2011/03/N/ST2/02652,
2013/08/M/ST2/00323,
and by the Foundation for Polish Science through the MPD programme and the project HOMING PLUS BIS/2011-4/3.
\section{Bibliography}
\nocite{*}
\bibliographystyle{elsarticle-num}
\bibliography{martin}

\end{document}